\documentclass{article}

\usepackage{arxiv}

\usepackage[utf8]{inputenc} 
\usepackage[T1]{fontenc}    
\usepackage{hyperref}       
\usepackage{url}            
\usepackage{booktabs}       
\usepackage{amsfonts}       
\usepackage{nicefrac}       
\usepackage{microtype}      
\usepackage{lipsum}		
\usepackage{graphicx}
\usepackage{natbib}
\usepackage{doi}

\usepackage{xcolor}
\usepackage{amsmath}
\usepackage{graphicx}
\usepackage{multirow}
\usepackage{multicol}
\usepackage{threeparttable}
\usepackage{subfigure}
\usepackage{booktabs}
\usepackage{amssymb}
\usepackage{makecell}

\title{LiteGEM: Lite Geometry Enhanced Molecular Representation Learning for Quantum Property Prediction}

\author{
 Shanzhuo Zhang\thanks{Equal contribution.}, Lihang Liu\footnotemark[1]  \\
 \textbf{Sheng Gao\footnotemark[1], Donglong He\footnotemark[1], Xiaomin Fang} \\
 PaddleHelix Team, Baidu Inc. \\
 \texttt{\{zhangshanzhuo, liulihang\}@baidu.com} \\
 \texttt{\{gaosheng06, hedonglong\}@baidu.com} \\
 \texttt{fangxiaomin01@baidu.com}
 \And
 Weibin Li\footnotemark[1], Zhengjie Huang\footnotemark[1] \\
 \textbf{Weiyue Su, Wenjin Wang} \\
 PGL Team, Baidu Inc.\\
 \texttt{\{liweibin02, huangzhengjie\}@baidu.com} \\
\texttt{\{suweiyue, wangwenjin02\}@baidu.com}
}
\date{}

\renewcommand{\headeright}{Technical Report}
\renewcommand{\undertitle}{Solution to KDD Cup 2021 PCQM4M-LSC challenge}
\renewcommand{\shorttitle}{LiteGEM}
\newcommand{\bm}{\mathbf{m}}
\newcommand{\h}{\mathbf{h}}
\newcommand{\ba}{\mathbf{a}}
\newcommand{\e}{\mathbf{e}}
\newcommand{\x}{\mathbf{x}}

\hypersetup{
pdftitle={A template for the arxiv style},
pdfsubject={q-bio.NC, q-bio.QM},
pdfauthor={David S.~Hippocampus, Elias D.~Striatum},
pdfkeywords={First keyword, Second keyword, More},
}

\begin{document}
\maketitle

 \begin{abstract}
In this report, we (SuperHelix team) present our solution to KDD Cup 2021-PCQM4M-LSC, a large-scale quantum chemistry dataset on predicting HOMO-LUMO gap of molecules. Our solution, {\bf Lite} {\bf G}eometry {\bf E}nhanced {\bf M}olecular representation learning (LiteGEM) achieves a mean absolute error (MAE) of 0.1204 on the test set with the help of deep graph neural networks and various self-supervised learning tasks. The code of the framework can be found in https://github.com/PaddlePaddle/PaddleHelix/tree/dev/competition/kddcup2021-PCQM4M-LSC/.
 \end{abstract}


\section{Introduction}
Molecular property prediction has been widely considered as one of the most critical tasks in computational drug and materials discovery since many methods rely on predicted molecular properties to evaluate, select and generate molecules. With the development of deep neural networks (DNNs), molecular representation learning exhibits a great advantage over feature engineering-based methods, which has attracted increasing research attention to tackling the molecular property prediction problem.
 
Inspired by our previous work (\cite{fang2021chemrlgem}), we propose using {\bf Lite} {\bf G}eometry {\bf E}nhanced {\bf M}olecular representation learning (LiteGEM) for the quantum property prediction: HOMO-LUMO gap (\cite{nakata_pubchemqc_2017}). We add the word ``Lite'' due to the fact that our original GEM model requires 3D geometry information as the input features, which is absent in this dataset. However, the self-supervised learning strategies proposed in GEM can still be incorporated into LiteGEM and boost the performance.

The report is organized as follows. We briefly introduce graph neural nets and message passing in Section~\ref{section:pre}. In Section~\ref{section:methods}, we present the main architecture of our model LiteGEM and its components. Experiment details such as choices of hyper-parameters and performance of ensemble can be found in Section~\ref{sec:experiments}.

\section{Preliminaries}
\label{section:pre}
A molecule consists of atoms, and the neighboring atoms are connected by the chemical bonds, which can be naturally represented by a graph $G=(\mathcal{V},\mathcal{E})$, where $\mathcal{V}$ is a node set and $\mathcal{E}$ is an edge set. An atom in the molecule is regarded as a node $v\in \mathcal{V}$ and a chemical bond in the molecule is regarded as an edge $(u, v) \in \mathcal{E}$ connecting atoms $u$ and $v$.

Graph neural networks (GNNs) can be seen as message passing
neural networks (\cite{DBLP:conf/icml/GilmerSRVD17}), which are useful for predicting molecular properties. Following the definitions of the previous GNNs (\cite{DBLP:conf/iclr/XuHLJ19}), the features of a node $v$ are represented by $\x_v$ and the features of an edge $(u, v)$ are represented by $\e_{uv}$. Taking node features, edge features and the graph structure as inputs, a GNN learns the representation vectors of the nodes and the entire graph, where the representation vector of a node $v$ is denoted by $\h_v$ and the representation vector of the entire graph is denoted by $\h_G$. A GNN iteratively updates a node's representation vector by aggregating the messages from the node's neighbors. Given a node $v$, its representation vector $\h_v^{(k)}$ at the $k$-th iteration is formalized by
\begin{equation}
    \begin{split}
        \ba_v^{(k)} &= \textit{AGGREGATE}^{(k)}(\{ (\h_v^{(k-1)}, \h_u^{(k-1)}, x_{uv} | u \in \mathcal{N}(v)\}), \\
        \h_v^{(k)} &= \textit{COMBINE}^{(k)}(\h_v^{(k-1)}, \ba_v^{(k)}).
    \end{split}
    \label{eq:gnn}
\end{equation}
where $\mathcal{N}(v)$ is the set of neighbors of node $v$, $\textit{AGGREGATE}^{(k)}$ is the aggregation function for aggregating messages from a node's neighborhood, and $\textit{COMBINE}^{(k)}$ is the update function for updating the node representation. We initialize $\h_v^{(0)}$ by the feature vector of node $v$, i.e., $\h_v^{(0)}=\x_v$.

$\textit{READOUT}$ function is introduced to integrate the nodes' representation vectors at the final iteration to gain the graph's representation vector $h_G$, which is formalized as
\begin{equation}
    \h_G=\textit{READOUT}({\h_v^{(K)}|v \in \mathcal{V}}),
\end{equation}
where $K$ is the number of iterations. In most cases, $\textit{READOUT}$ is a permutation invariant pooling function, such as summation and maximization.
The graph's representation vector $\h_G$ can then be used for downstream task predictions.




 
 \section{The LiteGEM Framework}\label{section:methods}
We introduce our solution in this section, including the GNN architectures, self-supervised learning strategies, and feature engineering tricks.

\subsection{LiteGEMConv}
Unlike Convolutional Neural Networks (CNNs), which are able to take advantage of stacking very deep layers, Graph Convolutional Networks (GCNs) suffer from vanishing gradients, over-smoothing, and over-fitting issues when going deeper. To encode the whole molecular structures, motivated by DeeperGCN from \cite{li2020deepergcn}, we propose a message-passing strategy for graph convolution: LiteGEMConv, which is formalized as 
\begin{align}
    \label{eq:msg}
    \bm_{vu}^{(k)} &= \textit{MLP} \left(\h_v^{(k-1)}\|\h_u^{(k-1)}\|\e_{vu} \right), \\
    \ba_v^{(k)} & = \textit{SoftMax\_Agg}_{\beta} \left( \left\lbrace  \bm_{vu}^{(k)}|u \in \mathcal{N}(v)\right\rbrace \right),\\
    \h_v^{(k)} &  = \textit{Linear} \left( \h_v^{(k-1)} + \ba_v^{(k)} \right),
\end{align}
where $\|$ denotes concatenation of vectors, $\textit{MLP}$ denotes a 2-layer Multi Layer Perceptron (MLP) with SiLU as the activation function and \textit{Linear} denotes the Linear layer. $\textit{SoftMax\_Agg}_{\beta}$ (\cite{li2020deepergcn}) function is used as our aggregation, which is defined as:

\begin{align}
\label{eq:softmax_aggr}
\textit{SoftMax\_Agg}_{\beta}(\cdot) = \sum_{u \in \mathcal{N}(v)} \frac{\exp \left(\beta \mathbf{m}_{v u}\right)}{\sum_{i \in \mathcal{N}(v)} \exp \left(\beta \mathbf{m}_{v i}\right)},
\end{align}

where $\mathbf{m}_{vu} \in \mathbb{R}^{D}$ is the given message set $\{ \mathbf{m}_{vu} | u \in \mathcal{N}(v) \}$\, and $\beta$ is the temperature controlling the smoothness of the distribution. Note that we also try to replace $\textit{SoftMax\_Agg}_{\beta}$ with a simple summation aggregation and observe no degradation in the performance, but we still keep it.

Overall, the LiteGEM consists of 11 LiteGEMConv layers with virtual node representations (\cite{gilmer2017neural}) added at each layer. Node features $\h_v$ initialized as $\x_v$ are updated by LiteGEMConv layer-by-layer, and the output is denoted as $\h_v^{(K)}$. The graph level representation $\h_G$ is obtained via mean pooling over the node representations of all nodes and the final output as the prediction of HOMO-LUMO gap is produced by another MLP as:
\begin{equation*}
    y_G = \text{MLP}_{G}(\h_G).
\end{equation*}
Here $\text{MLP}_{G}$ is a 2- or 3-layer MLP with dropout, batch normalization and SiLU activation. 



\subsection{Auxiliary and Pre-training Tasks}
Self-supervised learning methods have proven their effectiveness in computer vision and natural language tasks when the number of labeled data is insufficient. To further enhance our model performance, we adopt both auxiliary tasks and pre-training tasks in a self-supervised fashion for graph neural networks. We follow our previous work (\cite{fang2021chemrlgem}) and apply topology-level and geometry-level learning tasks as auxiliary and pre-training tasks.

\subsubsection{Geometry-level: Bond Length \& Bond Angle Prediction}

According to the Hohenberg–Kohn theorems in DFT, the ground-state electron density determines all ground-state properties of a molecule, i.e., the ground-state conformation determines the HOMO-LUMO gap of a molecule. On the other hand, as proved in our previous work ChemRL-GEM (\cite{fang2021chemrlgem}), by incorporating accurate 3D conformation in the QM9 dataset, we can achieve significant improvement in predicting various quantum properties of molecules. However, unlike ChemRL-GEM, due to the inference time limitation of this task, it is nearly impossible to generate accurate 3D conformation using DFT during inference.

To solve this dilemma, we adopt geometry-level self-supervised learning tasks of GEM only in the training procedure. More concretely, we utilize the bond length and bond angle prediction task of GEM as both the pre-training and auxiliary task for training LiteGEM. The bond lengths and the bond angles are the most important molecular geometrical parameters. The bond length is the distance between two joint atoms in a molecule, reflecting the bond strength between the atoms, while the bond angle is the angle connecting two consecutive bonds, including three atoms, describing the local spatial structure of a molecule.

\subsubsection{Topology-level: Context Prediction}
In context prediction, we use GNN to predict graph sub-structures (\cite{mikolov2013distributed}, \cite{rubenstein1965contextual}) using node-level representations. The intuition is that we want to enhance the ability of GNN to encode the neighborhood graph structure for each node. In order to represent the neighborhood of node $u$ in a compact way, we design rules to map the neighborhood into a \textit{context string}. Such string then contains all nodes and edges information that are at most 2 hops away from $u$ in the graph. Specifically, the context string contains the following sub-strings:
\begin{itemize}
    \item $\textit{str}(u)$;
    \item $\textit{str}(\textit{sort}(\mathcal{N}(u)))$: the 1-hop neighboring nodes of $u$;
    \item $\textit{str}(\textit{sort}(\{e_{uv}: v \in \mathcal{N}(u)\}))$: the 1-hop neighboring edges of $u$;
    \item $\textit{str}(\textit{sort}(\{w: w \in \mathcal{N}(v), v \in \mathcal{N}(u)\}))$: the 2-hop neighboring nodes of $u$;
    \item $\textit{str}(\textit{sort}(\{e_{vw}: w \in \mathcal{N}(v), v \in \mathcal{N}(u)\}))$: the 2-hop neighboring edges of $u$;
\end{itemize}
where we use $\mathcal{N}(v)$ to denote the neighboring nodes of node $u$, the function \textit{sort} is used to convert the set into a sorted array and the function \textit{str} is used to convert an array into a compact string. The above 5 sub-strings can then can concatenated to construct the context string. Note that we use the atomic number as the information for the nodes and the bond type for the edges. 
\par Upon obtaining the context string, we hash it into an integer between 0 and 5000 for each node. Then we construct a node-level multi-class classification task to predict these integers by adding a Multi-layer Perceptron Layer (MLP) after the node representations. The MLP we used is composed of 2 linear layers with Batch Normalization and SiLU activation. The only hidden layer has a size equal to half of the input dimension, and the output layer has a size of 5000. This fore-mentioned classification task is then used for both pre-training and auxiliary tasks. 

\subsection{Smoothing}

To further improve the performance, inspired by APPNP algorithm (\cite{klicpera2018predict}), we integrate a smoothing strategy after the last LiteGEMConv block. Our smoothing strategy utilizes the relationship between graph convolutional networks (GCN) and PageRank\cite{page1999pagerank} to derive an improved propagation scheme based on personalized PageRank. It achieves linear computational complexity by approximating topic-sensitive PageRank via power iteration. The process is defined as:
\begin{align}
	\begin{aligned} 
		\boldsymbol{Z}^{(0)} &=\mathbf{M}^{L}, \\ 
		\boldsymbol{Z}^{(k+1)} &=(1-\alpha) \widehat{\widetilde{\boldsymbol{A}}} \boldsymbol{Z}^{(k)}+\alpha \boldsymbol{M}^{L}. \\
	\end{aligned}
\end{align}
Here, $\mathbf{M}^{L} \in \mathbb{R}^{N \times D}$, is the node representations $\h_v^{(K)}$ produced by the DeeperGCN. $\widehat{\widetilde{\boldsymbol{A}}}$ is the adjacency matrix with self-loops and normalization. Note that when $\alpha=0$, the smoothing layer degenerates into an identity mapping.

\subsection{Feature Selection}

We extend the standard node and edge features provided by the OGB-PCQM4M dataset (\cite{doi:10.1021/acs.jcim.0c01224}, \cite{Chen_2019}, \cite{Kearnes_2016}). To be specific, 
we incorporate the \emph{atom type, chirality, degree, formal charge, number of hydrogen atoms connected, radical electron, hybridization, ring size, van
der Waals radius, the valence of out shell, partial charge} as atom features, and \emph{aromatic, if in ring, bond type, bond stereo type and conjugated } as bond features.

\section{Experiments}
	\label{sec:experiments}

\subsection{Implementation Details}

\begin{table}[]
\caption{Hyper-parameters}
\centering
\begin{tabular}{cl}
\toprule
  lr &
  [1e-4, 2e-4, 5e-4, 1e-3] \\

  dropout &
  [0.2, 0.3] \\

  activation &
  [SiLU, ReLU, GeLU, SoftPlus] \\

  $\alpha$ &
  [0, 0.2, 0.5, 0.8] \\

  $t$ &
  [1, 0, -1] \\

  batch size &
  256 \\

  embed dimension &
  1024 \\

  $w_{aux}$ &
  [0, 0.1, 0.2] \\ 
\bottomrule
\end{tabular}

\label{tab:hyper_param}
\end{table}

LiteGEM is implemented in PaddlePaddle\footnote{https://github.com/paddlepaddle/paddle} and PGL\footnote{https://github.com/PaddlePaddle/PGL}.
The total number of parameters is about 74 million for a single model. Other hyper-parameters for training the model are listed in Table \ref{tab:hyper_param}. The training time is about 100 minutes per epoch on an Nvidia Tesla V100. The model usually needs about 100 epochs to converge though we set the total number of epochs to 150. To fully make use of the validation set, we adopt a 2-fold cross-validation scheme. 90\% of the original validation set is randomly split into 2 cross-validation sets, and the last 10\% is left to verify strategies of the ensemble.
We train 5 variants of LiteGEM and save 164 checkpoints for further model ensemble (see Table \ref{tab:model_variants}).

The whole training program is preceded by a pre-training stage, where the model is only trained with the self-supervised tasks for 10 epochs.
After the pre-training, we combine the auxiliary tasks together with the HOMO-LUMO-gap task. To balance the trade-off between the auxiliary tasks and the HOMO-LUMO-gap task, we use a pre-chosen discount factor: the loss of auxiliary tasks is down-weighted by 1/2 at the 20th epoch and completely removed after the 30th epoch. The initial weight for auxiliary tasks ($w_{aux}$) is shown in Table \ref{tab:hyper_param}.

The 3D molecular conformation used in the bond length and bond angle tasks is generated by DFT using PySCF (\cite{sun_recent_2020}). We mimic the DFT process used for calculating the original dataset (\cite{nakata_pubchemqc_2017}) with various simplification to reduce resource consumption and generate 3D molecular conformation for all the 3 million training SMILES. Due to the simplification, only 20\% of DFT results are comparable with the original dataset and used for the auxiliary and pre-training tasks. The calculation for each molecule takes about 20 core minutes.

\subsection{Performance of Ensemble}

Due to the limit of inference time, we need to control the size of ensembles. To achieve this, we first drop more than half the model checkpoints by the coefficients obtained from a Huber Regressor. Specifically, checkpoints with an absolute coefficient less than 0.01 are deemed futile and dropped. All the 164 model checkpoints and the 73 checkpoints for the ensemble are listed in Table \ref{tab:model_variants}. Original models are regressed on the 2 cross-validation sets separately. 

The predictions of remained checkpoints are then sorted for each molecule, and the largest and smallest 20\% predictions are dropped. Remained predictions are regressed on another Huber Regressor. We use this regressor to produce the final prediction on the test set.
 
Without ensemble, the best single model achieved an MAE of $\sim$0.1235 on the last 10\% validation set, and the ensemble further decreased the MAE to 0.1141. On the whole test set, the MAE of our final ensemble model is 0.1204.

\begin{table}[]
\caption{Number of model checkpoints before and after ensemble}
\label{tab:model_variants}
\centering
\begin{tabular}{cccc|cc}
\toprule
\multicolumn{3}{c}{Auxiliary and Pre-training Tasks} &  &  &  \\ \cmidrule(r){1-3}
ContextPred & BondLength & BondAngle & Smoothing & \# Ckpts & \# Ensemble Ckpts \\ \midrule
- & - & - & Yes & 36 & 9 \\
- & - & - & - & 40 & 15 \\
Yes & - & - & - & 40 & 20 \\
Yes & Yes & - & - & 24 & 15 \\
Yes & Yes & Yes & - & 24 & 14 \\ \midrule
Sum &  &  &  & 164 & 73 \\ \bottomrule
\end{tabular}
\end{table}

\subsection{Inference Time}

Taking raw SMILES as input, the total inference time for all the 377,423 test molecules is about 10.5 hours on an Nvidia Tesla P40. To be specific, it takes 0.5 hour to generate graph features, 10 hours in inference by all the 73 ensemble models (8 minutes each). The time consumed by the Huber Regressor for the ensemble is negligible.

\subsection{Performance on 8-fold Cross-validation Sets}

Although not part of our submitted solution, we further test our model with only the context prediction as the auxiliary task on 8-fold cross-validation sets to make a comparison with the 1st-place winner solution proposed by \cite{ying2021place}. Our model achieves lower Valid MAE on all the 8 folds of data (see Table \ref{tab:cross_valid}), demonstrating the superiority of our architecture. 

\begin{table}[]
\caption{Valid MAE of LiteGEM on 8-fold cross-validation sets}
\label{tab:cross_valid}
\centering
\centering
    \begin{tabular}{ccc}
    \toprule
        Fold & \makecell[c]{LiteGEM\\w/ ContextPred} & Graphormer \\ \midrule
        0 & 0.0949 & 0.0970 \\ 
        1 & 0.0946 & 0.0971 \\ 
        2 & 0.0949 & 0.0970 \\ 
        3 & 0.0946 & 0.0965 \\ 
        4 & 0.0950 & 0.0968 \\ 
        5 & 0.0945 & 0.0967 \\ 
        6 & 0.0958 & 0.0965 \\ 
        7 & 0.0950 & 0.0969 \\ \bottomrule
    \end{tabular}
\end{table}

\section{Conclusions and Future Work}
In this challenge, our model LiteGEM achieves {\bf test MAE of 0.1204} on PCQM4M with the help of deep graph neural networks and self-supervised learning strategies. For future work, we will continue to study how to effectively incorporate knowledge from quantum mechanics. In the meantime, we sincerely thank the OGB team for their consistent hard work and effort in maintaining the contest and pushing the developments of graph neural networks.

\bibliographystyle{unsrtnat}
\bibliography{references}

\end{document}